\documentclass[12pt]{iopart}

\usepackage[
bookmarks=true, colorlinks=true, unicode=true,
urlcolor=blue,linkcolor=blue, anchorcolor=blue,
citecolor=blue, menucolor=blue, filecolor=blue,
]{hyperref}
\usepackage{cite}
\usepackage{braket}
\usepackage{graphicx}
\usepackage{multirow}

\begin{document}
    \title[Sample space filling analysis for boson sampling validation]{Boson sampling validation approach by examining the sample space filling}
    \author{A A Mazanik and A N Rubtsov$^{*}$}
    \address{Russian Quantum Center, Skolkovo, Moscow 143025, Russia}
    \address{$^*$ Author to whom any correspondence should be addressed.}
    \ead{ar@rqc.ru}
    \begin{abstract}
    Achieving a quantum computational advantage regime, and thus providing evidence against the extended Church-Turing thesis, remains one of the key challenges of modern science. Boson sampling seems to be a very promising platform in this regard, but to be confident of attaining  the advantage regime, one must provide evidence of operating with a correct boson sampling distribution, rather than with a pathological classically simulatable one. This problem is often called the validation problem, and it poses a major challenge to demonstrating unambiguous quantum advantage. In this work, using the recently proposed wave function network approach, we study the sample space filling behavior with increasing the number of collected samples. We show that due to the intrinsic nature of the boson sampling wave function, its filling behavior can be computationally efficiently distinguished from classically simulated cases. Therefore, we propose a new validation protocol based on the sample space filling analysis and test it for problems of up to $20$ photons injected into a $400$-mode interferometer. Due to its simplicity and computational efficiency, it can be used among other protocols to validate future experiments to provide more convincing results.
    \end{abstract}
    
    \vspace{2pc}
    
    \section{Introduction}
    A widely held conjecture in computer science, called the extended Church-Turing thesis~\cite{2010_nielsen_QuantumComputationQuantum, 2006_kaye_IntroductionQuantumComputing}, states that any reasonable physical device can be efficiently~(i.e.,~in~polynomial~time) simulated on a classical computer. However, it seems that the classical paradigm is not powerful enough for efficient simulation of quantum systems among others. Moreover, some promising quantum algorithms have already been proposed, such as Shor's algorithm for integer factoring \cite{1997_shor_PolynomialTimeAlgorithmsPrime}, which are able to provide significant speed-up over classical computers and hence experimentally violate the extended Church-Turing thesis and demonstrate a quantum computational advantage. 
    A scalable implementation of Shor's algorithm capable of outperforming classical computers requires a large-scale universal quantum computer and is beyond current capabilities, but nevertheless, experimentally achieving the quantum advantage regime has become one of the most exciting challenges in modern science \cite{2019_arute_QuantumSupremacyUsing, 2022_madsen_QuantumComputationalAdvantage}.

\begin{figure}[h!]
		\centering
		\includegraphics[width = 0.7\textwidth]{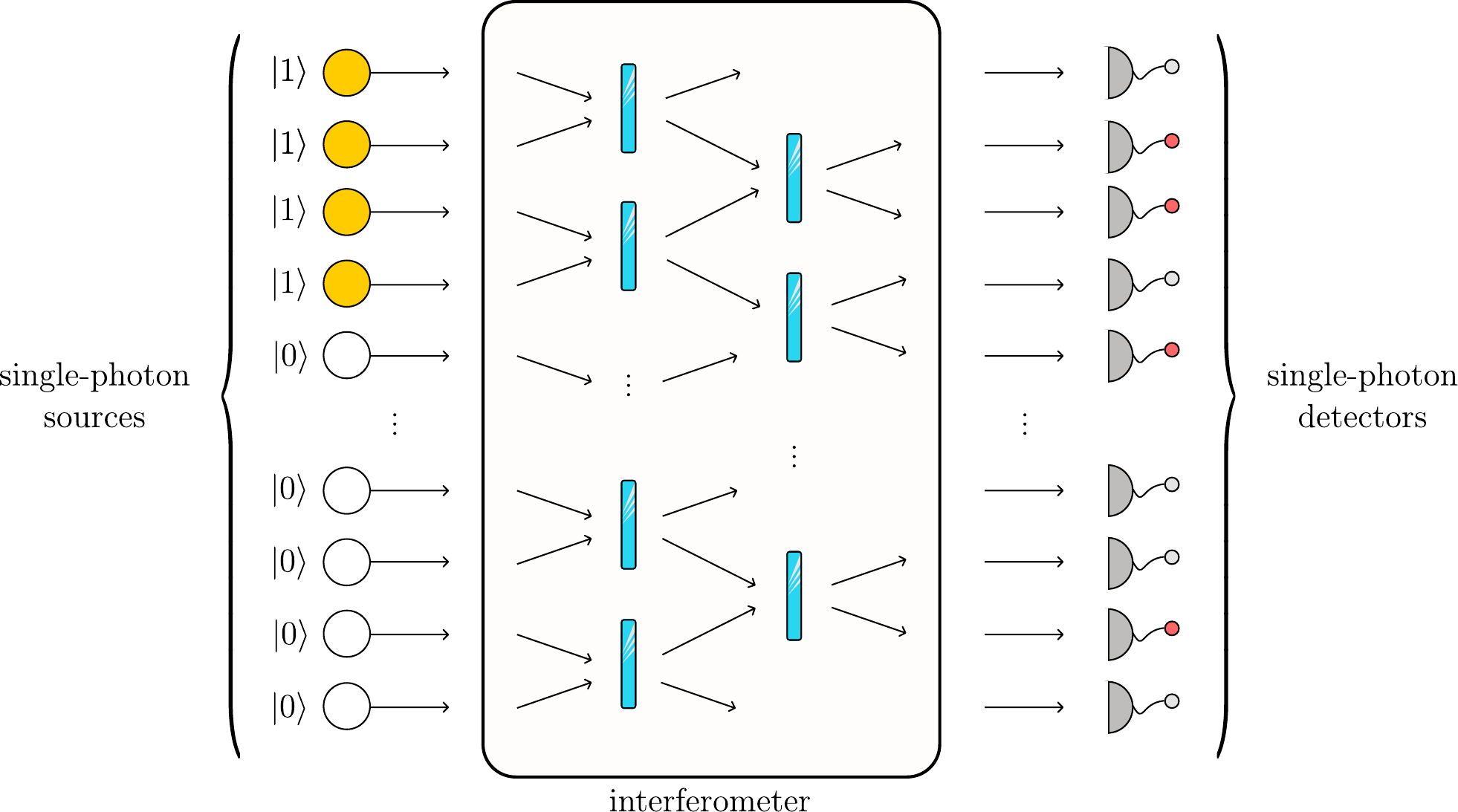}
		\caption{Boson sampling scheme. Single photon sources generate an initial Fock state, which is injected into an $m$-mode interferometer. It performs some unitary transformation on this input state, after which the output state is measured by single photon detectors. The red color schematically shows the detectors that have triggered.}
		\label{fig:bs_scheme}
    \end{figure}

    In this regard, Aaronson and Arkhipov proposed a computational problem they called boson sampling \cite{2011_aaronson_ComputationalComplexityLinear,2015_gard_IntroductionBosonSampling,2019_brod_PhotonicImplementationBoson}. Researchers provided compelling evidence that this problem is hard to solve on classical computers, while it can be solved efficiently using a photonic quantum hardware. This makes boson sampling a promising platform for demonstrating evidence against the extended Church-Turing thesis. 
    The original boson sampling scheme can generally be described as follows (see \fref{fig:bs_scheme}): single-photon sources generate an initial $m$-mode Fock state containing a total of $n$ photons, which is then injected into an interferometer performing some unitary transformation on the initial state. The output state is then measured by single-photon detectors. The result is a sample that is a string of $m$ non-negative integers summing to $n$, so the boson sampling problem is to sample from the distribution defined by a particular \mbox{output state.} 

    Since its initial proposal, many boson sampling experiments \cite{2013_crespi_IntegratedMultimodeInterferometers, 2013_tillmann_ExperimentalBosonSampling, 2013_spring_BosonSamplingPhotonic, 2013_broome_PhotonicBosonSampling,  2017_wang_HighefficiencyMultiphotonBoson, 2017_loredo_BosonSamplingSinglePhoton, 2017_he_TimeBinEncodedBosonSampling, 2018_wang_ScalableBosonSampling} have been carried out. To reduce experimental complexity and increase the size of photonic demonstrations, several other variants of boson sampling have been proposed, such as scattershot boson sampling \cite{2014_lund_BosonSamplingGaussian} or Gaussian boson sampling \cite{2017_hamilton_GaussianBosonSampling}, which are widely used in recent works \cite{2015_bentivegna_ExperimentalScattershotBoson, 2018_zhong_12PhotonEntanglementScalable, 2019_paesani_GenerationSamplingQuantum, 2020_zhong_QuantumComputationalAdvantage, 2022_madsen_QuantumComputationalAdvantage}. 
    
    The growing scale of these demonstrations has highlighted a new problem, often referred to as the validation problem. If there is an experimental setup that performs sampling procedure, its operation must be verified, i.e., it must be proven that this setup produces samples from the correct distribution. For small systems, as in the first experiments \cite{2013_crespi_IntegratedMultimodeInterferometers, 2013_broome_PhotonicBosonSampling, 2013_tillmann_ExperimentalBosonSampling, 2013_spring_BosonSamplingPhotonic}, it is possible to numerically calculate the correct output distribution and then compare it with that obtained experimentally. Yet, for larger systems, calculating the output distribution becomes computationally difficult on a classical computer, just as for experimental hardware, the number of samples required to correctly estimate the probability distribution increases significantly, so that a direct comparison of the two distributions becomes unfeasible. In this case, instead of precisely verifying the setup, one can at least reject for the experimentally obtained samples the hypotheses that they were sampled from several classically simulatable undesirable distributions, which therefore cannot provide a quantum advantage regime. In this formulation, the verification problem is often called validation, and it has been studied extensively over the past decade \cite{2014_aaronson_BosonsamplingFarUniform, 2014_carolan_ExperimentalVerificationQuantum, 2014_spagnolo_ExperimentalValidationPhotonic, 2016_liu_CertificationSchemeBoson, 2016_shchesnovich_UniversalityGeneralizedBunching, 2016_walschaers_StatisticalBenchmarkBosonSampling, 2018_giordani_ExperimentalStatisticalSignature, 2018_viggianiello_ExperimentalGeneralizedQuantum, 2019_agresti_PatternRecognitionTechniques, 2020_flamini_ValidatingMultiphotonQuantum, 2021_chabaud_EfficientVerificationBoson, 2022_giordani_CertificationGaussianBoson, 2023_iakovlev_BenchmarkingBosonSampler}. 

    Recently, a new approach based on network theory has been proposed \cite{2024_mendes-santos_WaveFunctionNetworkDescription} to analyse a complex wave function, given certain number of its measurements. As will be shown below, the boson sampling problem appears to be an excellent candidate for applying this methodology. It has already been discussed in this regard in \cite{2023_iakovlev_BenchmarkingBosonSampler}. 
    
    In this paper, we develop an approach to boson sampling validation based on constructing a wave function network and studying the dependence of the graph properties of the resulting network on the number of samples considered. We show that due to the intrinsic nature of the boson sampling problem, the dynamic of the sample space filling allows us to effectively distinguish the case of a real boson sampling from other classically simulated ones, without resorting to machine learning algorithms. We believe that due to its simplicity and direct generalisation to the other states of light, this method will help validate future boson sampling experiments. 
    
    The paper is organised as follows: \sref{sec:bs_problem} briefly introduces the formalism of the boson sampling problem and approaches to its validation. In \sref{sec:space-filling_analysis} we describe the construction of wave function networks and develop a new protocol for boson sampling validation. In \sref{sec:method_assessment} the protocol is tested for problems up to 20 photons in \mbox{a 400-mode interferometer.}  
    
    \section{Boson Sampling problem}
    \label{sec:bs_problem}
    \subsection{Boson Sampling formalism}
    Let us begin with a brief introduction to the formalism of the boson sampling problem. Consider the initial Fock state $\ket{\psi^{(in)}} = \ket{n_1,n_2,\dots,n_m}$ of $\sum_i n_i = n$ indistinguishable photons injected into the $m$-mode interferometer, specified by the operator $\hat{U}$ or the corresponding $m\times m$~Haar-random unitary matrix $\Lambda$. The state at the output can be written as follows \cite{2011_aaronson_ComputationalComplexityLinear, 2015_gard_IntroductionBosonSampling, 2019_brod_PhotonicImplementationBoson}:
    \begin{equation}
        \label{eq:wf_output}
        \eqalign{
        \ket{\psi^{(out)}} &= \hat{U}\ket{\psi^{(in)}} =\cr 
        &= \sum_S \gamma_S\ket{n^{(S)}_1,n^{(S)}_2,\dots,n^{(S)}_m}, \; S \in \Phi_{m,n},}   
    \end{equation}
    where the sum is over the set $\Phi_{m,n}$ of all possible ways to arrange $n$ identical photons in $m$~modes; $\gamma_S$ is the amplitude corresponding to the mode occupation \mbox{list $S = (n^{(S)}_1, \dots, n^{(S)}_m)$}. It therefore defines the probability distribution~$p_S=|\gamma_S|^2$ and sampling from this distribution is the crux of the boson sampling problem. 
    
    Given the transformation rule for boson creation operators $\mathbf{\hat{a}^\dagger} \rightarrow\Lambda^T\mathbf{\hat{a}^\dagger}$ \cite{2004_scheel_PermanentsLinearOptical},
     the amplitudes can be expressed using the permanents of $n\times n$ submatrices $\Lambda_S$ of $\Lambda$ as follows:
    \begin{equation}
        \gamma_S = \frac{\mathrm{perm}(\Lambda_S)}{\sqrt{n_1!\dots n_m! \times n^{(S)}_1!\dots n^{(S)}_m!}}.
    \end{equation}
    
    Since computing the permanents of complex matrices is known to be a \mbox{\#P-hard} problem, boson sampling turns out to be computationally inefficient for classical computers. Moreover, the approach to boson sampling based on simply computing the probabilities of each possible output $S \in \Phi_{m,n}$ requires computing $|\Phi_{m,n}| = C^{n}_{m+n-1}$ permanents, which grows very rapidly with the scale of the system. If we could efficiently compute at least the ratios of the permanents corresponding to different states, we could construct a Markov chain and use \mbox{Monte-Carlo} methods to simulate boson sampling device, but this is also impossible.

    \subsection{Boson Sampling validation}
    Some problems, such as integer factorisation, although difficult to solve, can still be verified efficiently. However, this is not the case for the boson sampling: there is no efficient algorithm for verifying its output. Despite this, any experimental setup that aims for a quantum advantage regime must have appropriate evidence of sampling from the correct distribution, rather than from the classically simulated one. The most straightforward way is to at least develop a special protocol that allows one to exclude sampling hypotheses from the most likely pathological distributions. Such protocols are often called validation protocols, and the problem of distinguishing the case of a real boson sampling from others is called validation \cite{2019_brod_PhotonicImplementationBoson}.

\begin{figure}[ht]
		\centering
		\includegraphics[width = 1\textwidth]{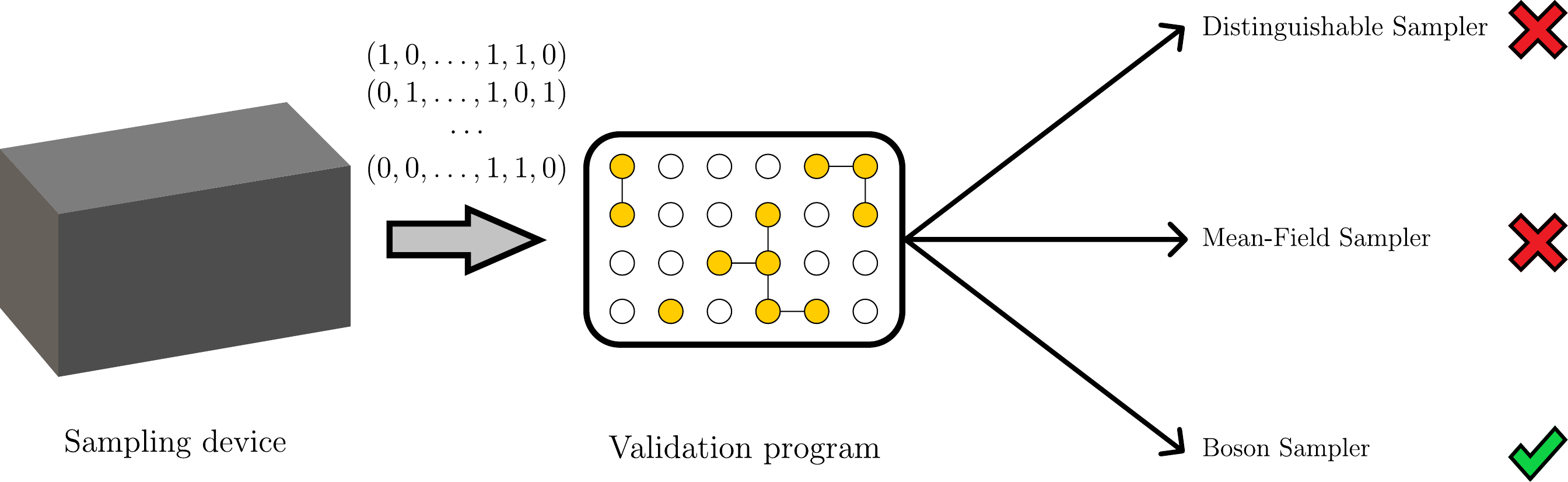}
		\caption{Schematic representation of the boson sampling validation protocol. Some unknown black-box sampling device produces samples that are fed to the input of some validation program. This validation program must provide a convincing conclusion about the similarity of the received data to several mock-up distributions.}
		\label{fig:validation}
    \end{figure}

    Schematically, any boson sampling validation protocol can be described as follows~(see \fref{fig:validation}): some unknown experimental setup or simulation program, which we will call a black box, produces samples from a set of possible outcomes, which are then fed to the input of a certain validation program. This program must, based on a very limited amount of data from the black box, provide a convincing conclusion about the similarity of the received data with that obtained from several plausible classically simulated distributions. If the black box data clearly differ from all classical cases considered, we can conclude that the samples were generated rather from the correct boson sampling distribution. Currently, there are three such classical distributions that are most widely considered: uniform \cite{2014_aaronson_BosonsamplingFarUniform}, distinguishable particles sampling \cite{2014_aaronson_BosonsamplingFarUniform, 2014_spagnolo_ExperimentalValidationPhotonic}, and mean-field sampling \cite{2014_tichy_StringentEfficientAssessment}.

    The validation problem historically began with the question of whether it is possible, for sufficiently large outcome spaces with a limited number of available samples, to distinguish the boson sampling distribution from a uniform one \cite{2013_gogolin_BosonSamplingLightSample}. However, rejecting the hypothesis of sampling from a uniform distribution is currently the most trivial task for any modern validation protocol. This is because it does not correlate at all with the system under consideration, so it is sufficient \cite{2014_aaronson_BosonsamplingFarUniform} to evaluate several single-particle observables reflecting certain properties of the interferometer to exclude a uniform distribution.
    
    More complex mock-up distribution assumes distinguishable particles at the input of the interferometer rather than indistinguishable bosons. It discards the effects of many-body quantum interference and can therefore be efficiently simulated classically~\cite{2014_aaronson_BosonsamplingFarUniform}. Distinguishable particles distribution can appear at the output of the boson sampling experimental setup due to its imperfections. Indeed, because of some internal unmeasured degree of freedom, such as frequency, polarisation, and arrival time, the photons can become partially distinguishable. Therefore, differentiating this case is of particular interest for the validation problem. Single-particle observables not more sufficient to distinguish the latter case from the real boson sampling. However, this can be ruled out by observing properties of multi-boson interference such as bunching or clouding \cite{2014_carolan_ExperimentalVerificationQuantum}. 
    
    To reproduce this clouding behavior, a mean-field boson sampling model was developed \cite{2014_tichy_StringentEfficientAssessment}. In the mean-field model, the multi-mode Fock state at the input semi-classically approximated by the macroscopically populated single-particle states with random phases. It is easily simulated using Monte-Carlo methods, so eliminating the hypothesis of sampling from the distribution determined by the mean-field sampler is now the most stringent test for boson sampling.

    An ideal validation protocol should not only be able to rule out the mock-up distributions described above, but should also not require resources that are exponential in the size of the problem \cite{2020_flamini_ValidatingMultiphotonQuantum}. That is, it should not include computing permanents of complex-valued matrices, nor should it require a large number of samples obtained from the black box. 

    The two most widely used validation protocols at present are statistical benchmarking \cite{2016_walschaers_StatisticalBenchmarkBosonSampling} and pattern recognition \cite{2019_agresti_PatternRecognitionTechniques}. The former requires computing a large set of correlators between the number of outgoing photons in each mode pair, while the latter uses special clustering algorithms to discover intrinsic patterns in the sample space. In this paper, we develop a new much simpler approach to boson sampling validation based on learning the filling of the sample space using wave function networks. It does not require any exponential resources or special machine learning methods and can be used together with the mentioned methods to provide a more convincing conclusion about the correct operation of the sampling device. With this work, we also want to highlight the power of network methods in learning complex many-body wave functions in quantum mechanical problems. 
    
    \section{Sample space filling analysis}
    \label{sec:space-filling_analysis}
    \subsection{Boson sampling wave function network}
    To demonstrate the best performance, any validation program should extract as much data as possible from the received set of samples. The usual way to do this is to analyse low-order correlation functions, but this discards much of the available information. Recently, Mendes-Santos~et~al.~\cite{2024_mendes-santos_WaveFunctionNetworkDescription} introduced wave function networks (WFNs), a mathematical tool for analysing complex many-body wave functions based on a very limited number of their measurements, called~``snapshots''. Boson sampling seems to be a natural candidate for applying this methodology. 

    Consider a set of snapshots of the wave function \eref{eq:wf_output} at the output of the interferometer $\{X_i\} = \{X_1, X_2, \dots, X_N\}$, where $N$ is the number of available snapshots, i.e. the number of samples received from the black box. We assume that there are no repetitions in the data set ($X_i \neq X_j, \forall \; i \neq j$). Each snapshot, labeled with index $i$, has the form:
    \begin{equation}
        X_i = (n_1^{(i)}, n_2^{(i)}, \dots, n_m^{(i)}) \in \Phi_{m,n},   
    \end{equation}
    where $n_p^{(i)}$ is the number of photons measured in the $p$-th mode. 
    
    In this paper, to compare distances between data points in the sample space we use the \mbox{$L^1$-metric:}~$d(X_i, X_j) = \sum_{p=1}^m{|n_p^{(i)} - n_p^{(j)}|}$. Once a metric is defined over the sample space, a criterion can be chosen to activate links between these data points based solely on the distances between them. This can be an activation distance $R$, such that if the distance between data points $d < R$, we activate the link. This distance can be selected manually, but a value that is too low or too high can lead to the loss of some information about the structure of the sample space, so it should be optimised to get the best result. 

    In this paper, we consider systems with the number of modes scaling quadratically with the number of photons $m = n^2$, and input states with the first $n$ modes occupied by single photons, the remaining modes left in vacuum. 
    
    \subsection{Dependence of network properties on the number of samples}
    In \cite{2023_iakovlev_BenchmarkingBosonSampler} it was shown that by comparing the degree distributions of wave function networks built on black-box data, the case of uniform sampling can be easily distinguished. It was also shown that, although differentiating between cases of distinguishable and indistinguishable bosons is a more difficult task, it can still be solved with high accuracy using machine learning techniques. In particular, the difference can be seen by calculating the mean $\mu$ and standard deviation $\sigma$ of the network degree distribution for several sampling procedures with different interferometer matrices $\Lambda$. This produces two clouds in the ($\mu,\sigma$)-plane, which can be well separated using various algorithms such as logistic regression.

    In general, there are two ways of forming such clouds. The first way corresponds to our knowledge of the transformation taking place inside the black box. In this case, clouds are formed by performing several sampling iterations on the same given interferometer matrix $\Lambda$. In contrast, the second way corresponds to an unknown black-box transformation, so the clouds are formed by performing sampling iterations on different Haar-random unitary matrices. In the latter case, the clouds are much bigger, so it is more difficult to validate. 

\begin{figure}[ht]
		\centering
		\includegraphics[width = 1\textwidth]{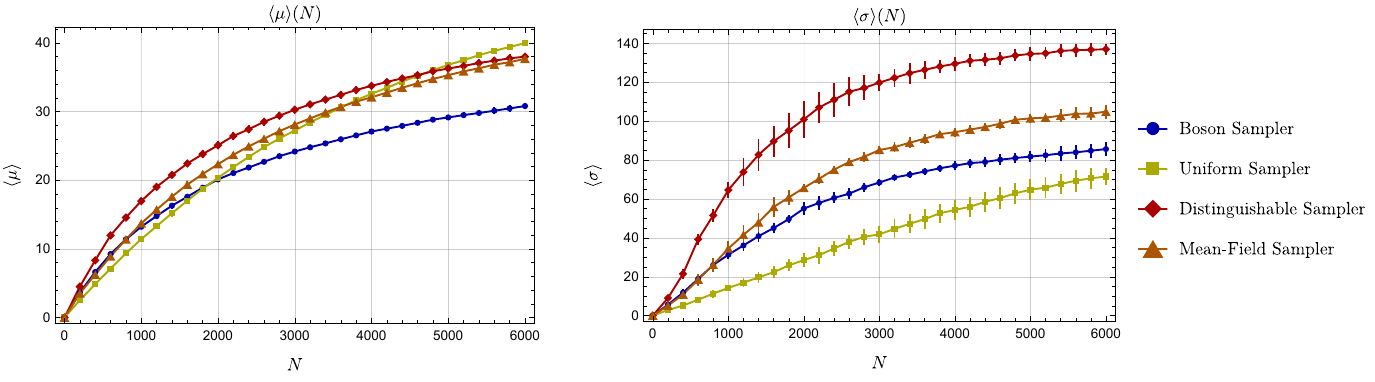}
		\caption{Dependence of the position of clouds in the plane ($\mu,\sigma$) on the number of samples considered $N$ for several sampling cases. The left image corresponds to the $\mu$-coordinate of the cloud, and the right one corresponds to the $\sigma$-coordinate. The vertical bars represent the variance of the distribution of cloud points along the corresponding axis and thus indicate the size of the cloud along this axis.}
		\label{fig:pos_4_14_6000}
    \end{figure}

    According to the definition of the activation radius, the mean of the degree distribution indicates the average number of a sample neighbors at distance $R$ in the sample space, and the standard deviation indicates the variance of this number of neighbors. Thus, including due to the finite size of the sample space, these quantities, and hence the average positions of clouds on the plane ($\mu,\sigma$), strongly depend on the number of samples $N$ being considered. \Fref{fig:pos_4_14_6000} shows an example of such a dependency for 4 photons in a 16-mode interferometer. Averaging was performed over several sampling iterations, with the same interferometer matrix $\Lambda$. Averaging over different interferometer matrices preserves the dependence of the cloud position, but increases~errors. 

    However, for systems large enough to be of practical interest to the validation problem, the number of available samples $N$ is always much smaller than the entire size of the sample space. This corresponds to the initial region of the curves shown in \fref{fig:pos_4_14_6000}, and as can be seen, the behavior of these curves in the considered region is noticeably different for each specific sampler that takes place. This allows us to develop a validation protocol based on the comparison of fitted parameters of the sample space filling curves.

    \subsection{Sample space filling analysis for boson sampling validation}

\begin{figure}[ht]
		\centering
		\includegraphics[width = 1\textwidth]{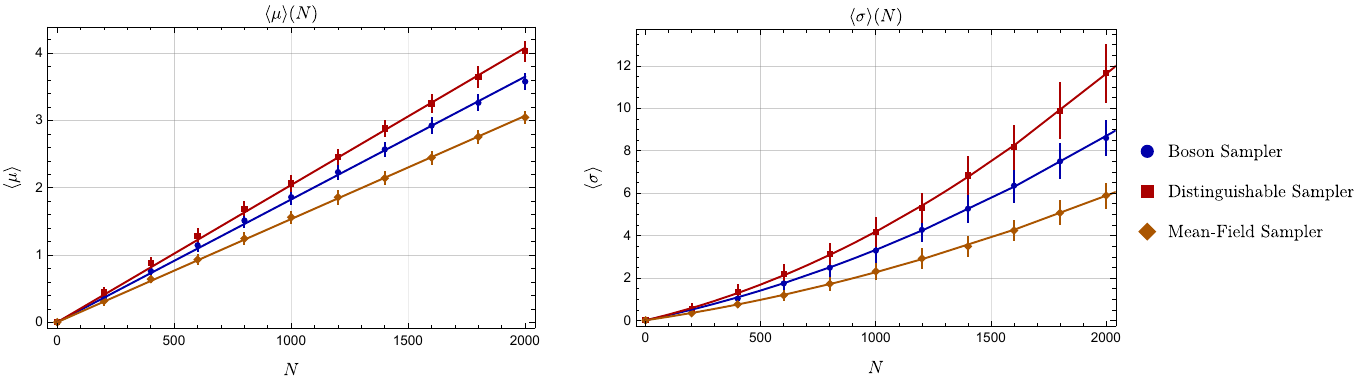}
		\caption{Cloud position in the ($\mu,\sigma$)-plane as a function of the number of collected samples up to $2,000$. The system size is $n = 5$, $m = 25$ with a fixed unitary transformation $\hat{U}$. The vertical bars indicate the cloud size, and the solid lines represent the fitted curves (see \eref{eq:fit} and \tref{tab:5ph}).}
		\label{fig:pos_5_25_}
    \end{figure}

    To illustrate our approach, consider a system of 5 photons injected into a 25-mode interferometer. There are $\Phi_{25,5} = C^5_{29} = 118,755$ possible outcomes. The cloud positions in the ($\mu,\sigma$)-plane are plotted against the number of samples in \fref{fig:pos_5_25_}. 
    Up to $2,000$ samples are considered, which is much less than the entire sampling space and in this region the dependence of the mean of the WFN distribution is linear for each sample instance, and the dependence of the standard deviation is quadratic. Thereby, these curves can be fitted by the low-order polynomials: 

    \begin{equation}
        \label{eq:fit}
        \cases{
            \langle\mu\rangle(N) = \alpha_{\mu}N, \\
            \langle\sigma\rangle(N) = \alpha_{\sigma}N + \beta_{\sigma}N^2.}
    \end{equation}

    This creates, for each given system under consideration, a correspondence between the sampling case occurring in the black box and the three numbers $\alpha_{\mu}, \alpha_{\sigma}$ and $\beta_{\sigma}$. These numbers do not depend on the number of samples received by the validation program, but characterise the sampling procedure itself. 
    For large systems it is practically impossible to go beyond the region where \eref{eq:fit} is applicable, so increasing number of samples obtained simply leads to an increase in the accuracy of the fitted values, as expected from a validation program \cite{2020_flamini_ValidatingMultiphotonQuantum}.

    The obtained fitted values for each sample instance are shown in \tref{tab:5ph}. The case of averaging over several transformations, i.e. over several interferometer matrices $\Lambda$, corresponds to less information about the black box and is more complex. However, as can be seen, despite the larger errors, all three sampling instances are still distinguishable. 
    \newpage
    \begin{table}[ht]
        \caption{\label{tab:5ph} Fitted values for cloud position dependence \eref{eq:fit}. Two cases considered: the case of averaging over sampling iterations for a known unitary transformation $\hat{U}$ and the case of averaging over different transformations.}
        \begin{indented}
        \lineup
        \item[]\begin{tabular}{@{}*{10}{l}}
        \br
        & \multicolumn{3}{c}{Boson Sampler} & \multicolumn{3}{c}{Distinguishable Sampler}\cr
        & $\alpha_{\mu} \times 10^{3}$ & $\alpha_{\sigma}\times 10^{3}$ & $\beta_{\sigma}\times 10^{6}$ & $\alpha_{\mu}\times 10^{3}$ & $\alpha_{\sigma}\times 10^{3}$ & $\beta_{\sigma}\times 10^{6}$\cr
        \mr
        single $\hat{U}$ & $1.80 \pm 0.03$ & $2.1 \pm 0.4$ & $1.09 \pm 0.32$ & $2.02 \pm 0.03$ & $2.1 \pm 0.7$ & $1.8 \pm 0.5$\cr
        $\0$diff $\hat{U}$ & $1.64 \pm 0.08$ & $1.8 \pm 0.9$ & $0.8 \pm 0.7$ & $1.88 \pm 0.11$ & $2.3 \pm 1.8$ & $1.2 \pm 1.3$\cr
        \br
        \end{tabular}
        \centering
        \item[]\begin{tabular}{@{}*{4}{l}}
        \br
        & \multicolumn{3}{c}{Mean-Field Sampler}\\
        & $\alpha_{\mu}\times 10^{3}$ & $\alpha_{\sigma}\times 10^{3}$ & $\beta_{\sigma}\times 10^{6}$\\
        \mr
        single $\hat{U}$ & $1.51 \pm 0.02$ & $1.60 \pm 0.32$ & $0.64 \pm 0.23$\\
        $\0$diff $\hat{U}$ & $1.43 \pm 0.06$ & $1.6 \pm 0.8$ & $0.6 \pm 0.6$\\
        \br
        \end{tabular}
        \end{indented}
    \end{table}
    
\begin{figure}[ht]
		\centering
		\includegraphics[width = 1\textwidth]{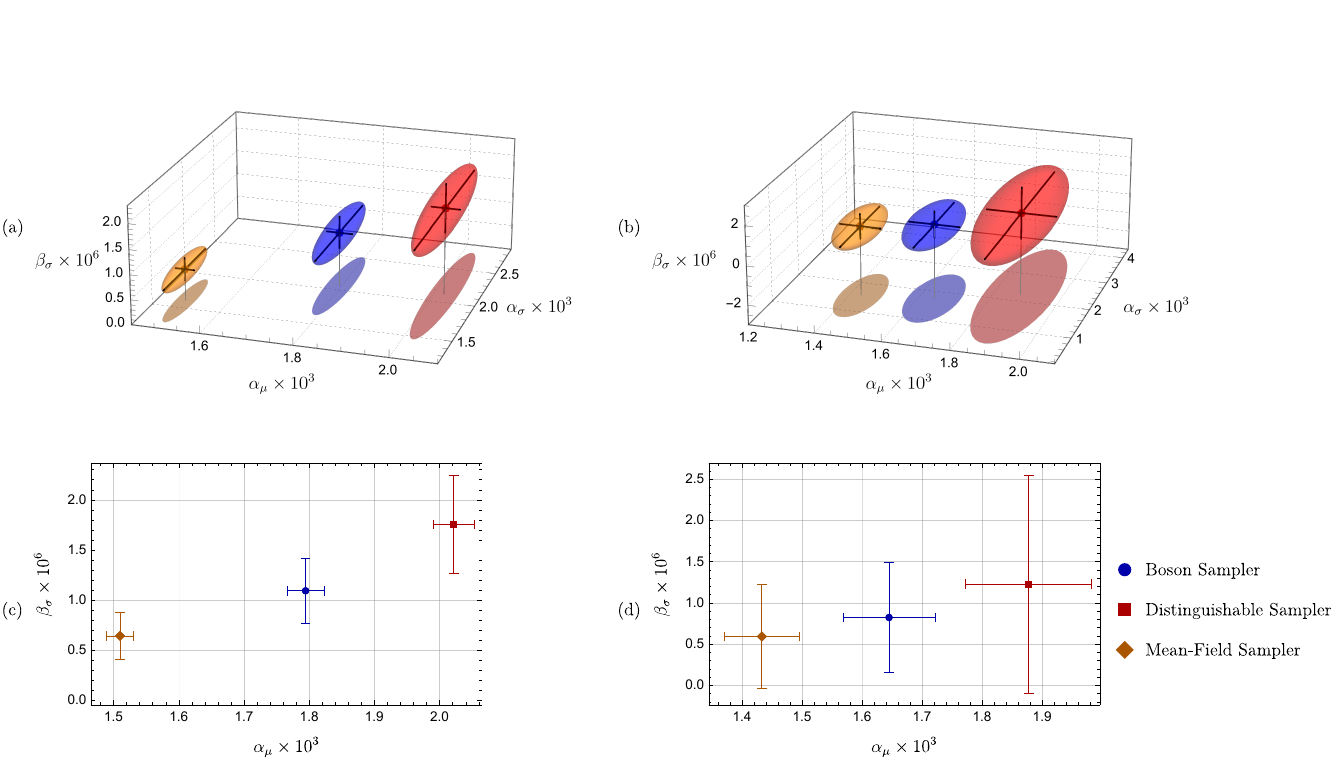}
		\caption{Three fitted parameters of cloud position dependence on the number of collected samples, together with ellipsoids corresponding to their errors: (a) averaged over sampling iterations with a fixed interferometer, (b) averaged over sampling iterations with several different interferometers. Vertical lines included as a visual aid to projections onto the $(\alpha_{\mu},\alpha_{\sigma})$ plane. (c, d): Projection of clouds onto the most informative $(\alpha_{\mu},\beta_{\sigma})$ plane for fixed interferometer and different interferometers, respectively. System size $n = 5$, $m = 25$; up to $2,000$ samples were collected.} 
		\label{fig:5ph_25m_3d}
    \end{figure}

    For greater clarity, it is better to depict the fitted values on a three-dimensional graph along with their errors. The corresponding graph is shown in \fref{fig:5ph_25m_3d}. In both cases, the error ellipsoids corresponding to the three distributions considered do not intersect, which indicates the sufficiency of the considered properties of WFN for solving the validation problem.  

    \newpage
    Thereby, we have demonstrated the possibility of distinguishing the case of a real boson sampling from the most commonly used mock up distributions by analysing the evolution of WFN properties with filling the sample space. The described method does not involve the calculation of permanents of complex-valued matrices, as well as  does not require a large number of samples obtained from the black box.   
    
    \section{Assessing the protocol on larger circuits}
    \label{sec:method_assessment}
    In the previous section, the approach was demonstrated for a relatively small circuit, however, it should also be tested for larger circuits, where the validation problem is most relevant. Indeed, for such circuits, very few samples are always available compared to the entire sampling space, and the filling curves described above can, in theory, become indistinguishable, rendering the protocol useless.

    Obtaining samples from the boson sampling distribution for large circuits is computationally expensive, and to overcome this problem we tested our approach using data from \cite{2017_neville_ClassicalBosonSampling}, which is openly published at the University of Bristol data repository. Since only sample sets from a few different samplers are available in the repository, these data represent a good model of the black box output and bring us closer to validating the real experimental setup.

\begin{figure}[ht]
		\centering
		\includegraphics[width = 1\textwidth]{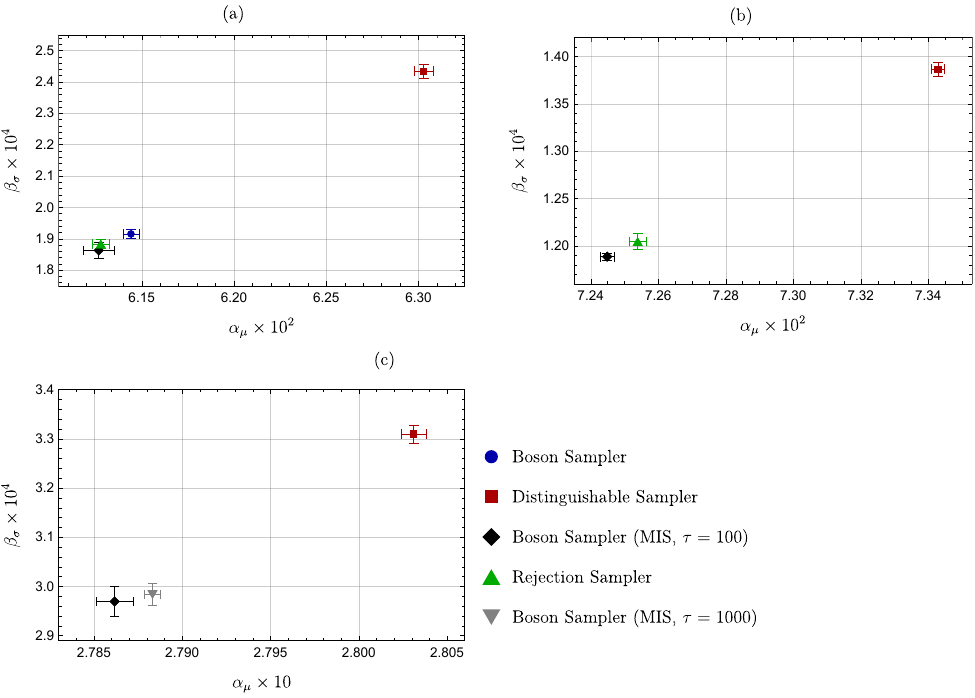}
		\caption{Two fitted parameters $\alpha_{\mu}$ and $\beta_{\sigma}$ of cloud position dependence on the number of collected samples together with their errors. Up to $18,000$ samples considered; the activation distances $R$ were optimised manually. (a): System size $n = 7$, $m = 49$; $R = 8$. (b): System size $n = 12$, $m = 144$; $R = 18$. (c): System size $n = 20$, $m = 400$; $R = 36$. Detailed information on the samplers presented is available in the supplementary material to \cite{2017_neville_ClassicalBosonSampling}.}
		\label{fig:larger_circuits}
    \end{figure}

    The systems of size $n = (7,\; 12,\; 20)$ and $m = (49,\; 144,\; 400)$, respectively, are considered. For the number of collected samples up to $18,000$ and for all sampling instances taking place, the fitting parameter $\alpha_{\sigma}$ corresponding to the linear part of the $\langle\sigma\rangle$-coordinate dependence on $N$ turns out to be zero within the error limits. Therefore, only two fitting parameters $\alpha_{\mu}$ and $\beta_{\sigma}$ remain, which allows us to plot them on a two-dimensional graph instead of a three-dimensional one. The resulting graphs are shown in \fref{fig:larger_circuits}.

    The samples from \cite{2017_neville_ClassicalBosonSampling} are obtained from a collision-free subspace, i.e. there is at most one photon in each output mode. Although this case is of particular difficulty for the validation problem \cite{2023_iakovlev_BenchmarkingBosonSampler}, as can be seen in \fref{fig:larger_circuits}, all cases that should represent a sample from the boson sampling distribution are well separated from the case of distinguishable particles. One can notice, that the rejection sampler and the two Monte-Carlo samplers are also slightly differ from the naive brute-force sampler, but there are still much closer to the real boson sampling than to a set-up with distinguishable particles.
    
    There are no data representing the mean-field sampler, and all data correspond to the simpler case of a single fixed unitary transformation $\hat{U}$. Nevertheless, the obtained results lead to the conclusion that the described approach is still applicable for systems of at least 20 photons injected into a 400-mode interferometer, which is quite acceptable for experimental needs.

    \section{Discussion}
    The validation problem remains one of the main challenges towards achieving an unambiguous quantum advantage regime and is related to distinguishing between multiple sampling cases. In this paper, we developed a protocol for boson sampling validation based on comparing the behavior of sample space filling for different sampling instances. We used wave function network approach to describe each current filling state. This allowed us to calculate the average number of neighbors $\langle\mu\rangle$ of a sample at some fixed distance $R$, as well as the variation $\langle\sigma\rangle$ in this number of neighbors. We found that the dependencies of these quantities on the number of collected samples have the same shape in the region of available sample numbers. Nevertheless, they have different fitting parameters, which allows us to distinguish between them. In contrast to a direct comparison of the positions of several clouds in the $(\mu,\sigma)$-plane, our fitting parameters do not depend on the specific number of collected samples and, for a given distance $R$, characterise the sampling procedure itself.

    Using numerical simulation data from \cite{2017_neville_ClassicalBosonSampling}, we tested our protocol for problem sizes of 7, 12 and 20 photons injected into a 49-, 144- and 400-mode interferometers, respectively. Although these data were obtained for a collision-free subspace, which is more difficult case to validate, our approach provides a robust distinction between the classically simulated case of distinguishable particles sampling and cases corresponding to a real boson sampling. Nevertheless, the simulation data used were obtained for a fixed interferometer matrix $\Lambda$, which, on the contrary, represents a simpler case consistent with our knowledge of the interferometer inside the black-box.

    The proposed approach does not require the calculation of any permanents of complex-valued matrices, and due to the adjustment of the activation distance $R$, it does not require a large number of collected samples compared to the entire size of the sample space, as well. The main computational complexity of our approach is related to the calculation of distances between all pairs of obtained samples. However, it is possible that modern methods of nearest neighbors search, widely used in unsupervised machine learning algorithms, should be applied here to optimise the calculation in the regime of a large number of modes and a large number of obtained samples.

    We leave it for further research whether it is better to use some other metric on the sample space to obtain more reliable results with fewer samples collected. There may be some other properties of the constructed graph that, in theory, can better capture the difference between the cases discussed. 
    It is also easy to extend the proposed approach to generalisations of the boson sampling problem dealing with Gaussian states of light. Perhaps there might be some special network structure for such cases as sampling with Gaussian, coherent or thermal states of light at the input, which would also be of particular interest. Therefore, we see that boson sampling provides a large platform for network-based research.

    \section{Online content}
    The code used to perform the calculations described in this paper is publicly available in the following GitHub repository: \\ \url{https://github.com/Alexandr-Mazanik/WFN-BosonSampling/tree/main}

    \section{Acknowledgments}
    This work was supported by Rosatom in the framework of the Roadmap for Quantum computing (Contract No. 868-1.3-15/15-2021 dated October 5).
    
    \section*{References}
    \bibliographystyle{iopart-num}
    \bibliography{QuantComputing}
    
\end{document}